**Title: Application of AI in Credit Risk Scoring for Small Business Loans: A case study on how AI-based random forest model improves a Delphi model outcome in the case of Azerbaijani SMEs.**

Author: Nigar Karimova

# Table of Contents

# Abstract

The research investigates how the application of a machine-learning random forest model improves the accuracy and precision of a Delphi model. The context of the research is Azerbaijani SMEs and the data for the study has been obtained from a financial institution which had gathered it from the enterprises (as there is no public data on local SMEs, it was not practical to verify the data independently).

The research used accuracy, precision, recall and F-1 scores for both models to compare them and run the algorithms in Python. The findings showed that accuracy, precision, recall and F-1 all improve considerably (from 0.69 to 0.83, from 0.65 to 0.81, from 0.56 to 0.77 and from 0.58 to 0.79, respectively).

The implications are that by applying AI models in credit risk modeling, financial institutions can improve the accuracy of identifying potential defaulters which would reduce their credit risk. In addition, an unfair rejection of credit access for SMEs would also go down having a significant contribution to an economic growth in the economy.

Finally, such ethical issues as transparency of algorithms and biases in historical data should be taken on board while making decisions based on AI algorithms in order to reduce mechanical dependence on algorithms that cannot be justified in practice.

# Introduction

The application of AI in a broad number of fields has gained momentum in recent years which has bolstered efficiency and effectiveness of organizational outcomes (Cao, 2022). Financial system is not an exception which has benefited from multiple contributions of AI and ML. Credit risk modeling is one of the areas which is prone to human errors (Cao, 2020). Additionally, manual work that goes into the calculation of default scores through assessment of the circumstances of each and every entry in the database is a cumbersome and time-consuming process (Li, et al., 2023). AI is, however, not only powerful for its completion of long processes in a relatively short period of time but also can capture complex relationships better compared to traditional models. Traditional models rely on the static nature of variables in the financial world which is not the case (Li, et al., 2023). A dynamic nature of variables must be incorporated into the analysis in order to make more informed decisions. Learning from the historical data, machine learning algorithms can provide a much more effective outcome in terms of accuracy (Farahani and Ghasemi, 2024). AI models do not only investigate internal factors such as performance measures of companies but also analyze macroeconomic and political (external factors) factors which can have a consequential impact on the performance of firms. SMEs are exposed to a great risk as changes in macroeconomic and political conditions in a particular country can be especially harsh on smaller firms (Li, et al., 2023). These benefits of the application of AI in financial analysis increased its use and the applications in credit scoring are prevalent as well. Nevertheless, the caveat is that there should be a large dataset relevant to a particular context in order for AI models to analyze and learn from for an efficient model.

The role of credit scoring and the integration of AI into the process for the successful operations of SMEs in emerging market economies are crucial (Popovych, 2022). Firstly, traditional scoring models rely on historical records and do not capture a wide variety of factors that can in combination contribute to the success or failure of SMEs. Therefore, more comprehensive approaches including the integration of machine learning into the process enable financial institutions to more accurately evaluate potential risks of SMEs and take into consideration variables which fluctuate with season and under the influence of general macroeconomic conditions (Popovych, 2022). Therefore, to appease unfair treatment of SMEs by financial institutions due to a lack of long historical data which often leads to either a rejection of loan application or high interest rates, AI in credit scoring holds a significant potential.

By evaluating possible AI models that can be applied in credit risk modeling, the purpose of this research is to introduce a AI-based model which improves the accuracy and precision of traditional models used in credit assessment. Random forests model is the selected algorithm for this purpose which adapts to real time data and incorporates non-linear complex relationships among variables to a significant extent (Zhang, et al., 2018). By using historical data to build simulations and learn from thousands of iterations, AI models also predict the variations in external variables and their impact on the performance of SMEs instead of focusing on limited historical data of the performance of SMEs themselves. The research aims to see a practical improvement in the accuracy and prediction of the model used traditionally for credit scoring after the application of random forest model. By providing the management of SMEs and financial institutions with valuable insights, the model will be crucial in terms of determining whether practical improvements can be obtained from the application of AI which leads to a greater accuracy in credit scoring. Moreover, the identification of relative weight of each of the risk-drivers will also allow SMEs to understand the most important risk categories they encounter.

Thesis statement: This study analyzes the effectiveness of AI algorithms in improving credit risk scoring models for SMEs by taking into consideration both advantages and ethical concerns. The research looks, in particular, into the case of Azerbaijani SMEs which have applied for bank loans in the last 5 years.

# Literature Review

## Previous Research on Credit Risk Scoring. A summary of existing models

The most fundamental of the available historical models for credit risk scoring is a regression analysis (Bolton, 2009). A particular type of regression called logistic regression is applied to predict the default of firms. Regression techniques involve the analysis of the impact of selected independent variables (characteristics of the loan applicant such as debt to equity, previous record of payments etc) and the dependent variable (default status) (Bensic, et al., 2005). Based on the historical data on the independent and dependent variables, a relationship between the two is identified allowing to predict defaults. Some limitations of these models should be highlighted. Firstly, logistic regression depends on linearity assumptions (Khemeais, 2016). The model assumes that there is a linear relationship between independent and dependent variables and this assumption is not always valid. Therefore, any deviations from linearity reduces the accuracy of logistic regression outcome (Bensic, et al., 2005). Furthermore, regression analysis uses historical data only which is again one of the limitations of this type of analysis for credit scoring as changes in the external environment cannot be captured by the analysis tool.

Next, discriminant analysis should be highlighted as a frequently used model for credit score modeling (Mvula, 2011). Discriminant analysis analyzes key characteristics of the companies studied and groups them into those likely to default and those which are in a better financial condition. The model has been in use for a long period of time but its assumptions are also restrictive as it relies on normal distribution of explanatory variables (Mylonakis, 2010). The model is also prone to have a distorted outcome under the condition of high volatility. In these cases, discriminant analysis performs poorly illustrating that it has a high accuracy under relatively stable conditions.

Finally, Altman's Z score model is a predictive model for default probability. The model incorporates financial ratios and using company fundamental information, it is possible to predict the likelihood of default (Panigrahi, 2019). A wide range of financial ratios go into the calculation of Z score and depending on the outcome score, the likelihood of default can be predicted. Some drawbacks of the model also exist as it has been mostly applied to large corporations and there can be volatility in emerging markets which cannot be reflected well by this model (Altman, et al., 2017).

In sum, different models have been applied thus far for credit score modeling but each has certain shortcomings.

## AI in Finance

AI has revolutionized finance and has been applied in credit score modeling as well extensively (Cao, 2022). AI in credit risk assessment has been used on multiple fronts. Firstly, the capability to analyze a vast amount of data in a short period of time has been one of the most attractive features of the application of AI. Additionally, the complexity of analyzed data is particularly impressive with AI as it can effectively divulge relationships between complex variables, capture non-linearity and provide an accurate outcome for predictions (Hilpisch, 2020). AI models also reduces noise in the data indicating that these models select the most important variables in the dataset.

AI also includes data which is unrelated to the direct activity of a borrower despite having a significant impact on the borrower default (Buchanan, 2019). These variables are external factors such as overall economic conditions in the economy, trends in the market, consumer behavior and even political risks. By capturing real-time data and disclosing its impact on risk factors have increased the prominence of AI in financial markets (Cao and Zhai, 2022). Deep learning models and neural networks are examples of these algorithms (machine learning) which have increased the effectiveness of credit scoring by financial institutions.

## Ethical concerns in AI

Depending on the algorithms and provided input data, ethical issues might be present in the use of AI for credit risk modeling. For example, existing discrimination in lending to a particular group of individuals can be reflected in AI solutions if these biases are a part of historical data used in training (Svetlova, 2022).

Regarding transparency, there is a literature on concerns about the algorithms used in deep learning models (Max, et al., 2021). These models are referred to as a black box and have raised concerns among consumers and other users who are not certain of the exact procedures the model uses and whether there are biases. The use of more explainable techniques such as LIME and SHAP has been advocated in order to ensure that AI models are more explainable (Bhattacharya, 2022).

Data use has serious implications. Therefore, relevant regulations have developed in this area. However, technology usually precedes regulations and with regards to the use of AI, the adherence to regulatory requirements has been rather ambiguous as the biggest developers of AI algorithms have operated mostly with a lack of transparency (Dumouchel, 2023). GDPR requirements have stated that usage of personal data must meet certain standards such as being based on unequivocal and informed consent. Nevertheless, there are certain concerns regarding whether all information used for AI data training and models is based on informed consent (Schuett, 2023). The issues of content and transparency have been present in regards to the use of AI and it can be argued that they are not resolved yet as AI is still in significant progress and similar issues are expected to emerge with time.

To summarize the literature review, models in credit scoring preceding AI use have been regression analysis, discriminant analysis and Altman's Z score model. These models have had serious drawbacks which mostly boil down to normality of data and linearity of relationship between variables. These features reduced the effectiveness of the mentioned models and increased preference for AI in the credit risk assessment process. Secondly, AI has found its way into credit risk assessment due to its ability to process a vast amount of data and capabilities regarding the analysis of non-linear relationships between variables. Credit scoring outcomes (such as precision and accuracy) have improved as a result of the use of AI in finance. Finally, ethicality remains an issue in AI credit score modeling because of the potential biases existing in the database used in the analysis and the way algorithms are built as there is a low level of transparency in the process.

## Methodology

### AI model used

The model used in the research is random forest model is a machine learning model/algorithm which allows researchers to enhance the accuracy of predictions which is suitable in this research as well (Van, et al., 2016). The power of the model lies with the fact that it incorporates multiple decision-trees and captures non-linear relationships between variables which is one of the shortcomings of traditional models. Considering that there are multiple dynamic risk factors impacting SMEs in Azerbaijan and in general, the model provides useful insights into the complex nature of relationships between variables.

Several steps are taken in the model to arrive at a final outcome (prediction). Firstly, the model takes sample data from the original data and creates training datasets. Randomness of the selection of data points from the original data and replacement contributes to the robustness of the model. This step is bootstrap sampling.

Next, for each training dataset, a decision-tree is built. Decision trees incorporate a number of key variables such as revenue growth, leverage and so on. Based on the trained data relevant to each dataset, a decision is made by the model which is the most suitable (Van, et al,. 2016).

Decisions for each decision tree is made individually but aggregated for a final result. The final result reflects the probability of default and tells whether the company will default or not.

The value of random forest models is high in the case of SME default projections because these models are capable of handling non-linear data and relationships. Oftentimes, relationships between variables cannot be analyzed accurately assuming linear relationships. Therefore, random forest models are a powerful tool in this regard.

Additionally, allocation of scores for the importance of each variable (in terms of which factors drive the company's default risk) plays a role in the analysis of SMEs. SME managers and creditors to these organizations can obtain insights from the model regarding which variables have contributed to the risk of default most.

Data sources

Data source for the research was a local bank in Azerbaijan which in consultation with SMEs allowed to access the data for SMEs used in the analysis. The same bank applied traditional credit scoring (Delphi method) to assess the risk of default of the analyzed SMEs in this study. 5-year data was collected for the research as the data for longer than 5-year horizon did not exist in the database (either because SMEs evaluated no longer had a relationship with the bank for more than 5 years or some of them stopped their operations altogether due to a default or being acquired by other firms).

The following historical data was analyzed as an input in the research:

-Revenue growth (historical revenue growth figures were taken into account). Revenue growth figures would be a useful starting point for the model to assess the future of the operations of SMEs. Furthermore, net profit margin of companies have also been selected as a variable as revenue might be misleading as a before costs measure.

-Cash flow variance: Cash flow is the most important metric in regards to credit assessment as cash flow health of the firm is the indicator of its ability to service its debt as well (Zhang, et al., 2018). Therefore, cash flow variability should be included in the calculation of default probability. Operating cash flow is the selected proxy in this research.

-Debt to equity ratio: This ratio illustrates the level of indebtedness (the proportion of debt in capital structure vs equity) and can be informative in terms of evaluation of the risks of companies. Excessive debt level can result in high risks as well preventing the firm from being able to pay off its debt. Therefore, it is crucial that debt to equity ratio is chosen as one of the variables in the model.

-Commodity price sensitivity: Commodity price sensitivity can be a crucial determinant of the risks faced by companies in particular small firms. Commodity prices on the world markets such as oil and gas, wheat and coffee can determine revenue level and profitability of SMEs. Therefore, based on the companies in the data, the most relevant commodities have been selected in order to measure the sensitivity of revenue of SMEs to the prices of these commodities. Oil price is the most pertinent due to the fact that Azerbaijan is an oil exporting country and its economic performance depends significantly on the price of oil. Therefore, a drop in the price of oil in the world markets can reduce profitability of companies in the country. Additionally, grain and cotton prices have been selected as a significant variable in regards to potentially explaining the revenue variability of Azerbaijani SMEs. These variables have been picked because SMEs in the sample are from agricultural sector also having an exposure to oil and gas prices. Sensitivity is measured by 5-year correlation coefficient between the prices of relevant commodities and revenue of SMEs.

-Market conditions: To measure the impact of market conditions, GDP growth rate has been integrated into the analysis as an input.

Model evaluation

Accuracy and precision of the model are metrics used for the evaluation of the model performance.

Ethical considerations

To ensure that there are no ethical issues in the model (biases), the data for the model has been selected so that it captures all available SME data provided instead of selecting companies from certain sectors or with certain characteristics (for example, poor performers vs good performers).

# Results

The results of the study have been presented in a comparative fashion. Firstly, the outcome of a Delphi model has been presented and discussed. Next, a random forest model implemented in the research as an improvement to the traditional method has been laid out. The comparison between key metrics facilitated an understanding of an improvement achieved as a result of the application of AI model.

It should be mentioned that variables (predictors) for each model are the same in the analysis. Thus, as discussed in methodology section, independent variables of the model are revenue growth, cash flow variance, leverage (debt to equity ratio), net profit margin, sensitivity to commodity prices (oil, cotton and grain have been included in the form of correlation coefficients between each of these commodity and revenue of SMEs) and default dummy (0 for no default and 1 for default that occurred).

Python code for the analysis can be found in Appendix 1.

Logistic regression has been applied (as one of the traditional methods of credit score modeling) as the Delphi model used in practice relied on a regression analysis according to the provider of the model (financial institution that applied this model to calculate credit scores of SMEs). Considering this fact, a logistic regression has been run (See Appendix 1 for code). Python codes have been run to calculate accuracy, precision, recall and F-1 scores.

To elaborate on the metrics, precision of the model refers to how often the prediction of the true positive by the model is correct. If there are both true and false predictions, the precision of the model is calculated by dividing the number of total correct positive predictions by the number of total predictions made (Provenzano, et al., 2020).

Precision is the accuracy with regards to the target class. However, accuracy of the model shows an overall accuracy of the model.

Furthermore, the ability of the model to find all of the objects in the target class is called a recall in machine learning which is also compared between two models (Provenzano, et al., 2020).

F-1 score, finally, is a balanced measure (harmonic mean) of precision and recall demonstrating an overall robustness of the model.

Next, Python code for a machine learning model (random forest) has been presented in Appendix 2 which is a step-by-step process for calculation of each metric (accuracy, precision, recall and F-1.

Table 1. Key performance indicators

| Performance metric | Delphi model | Random forest (AI model) |
|---|---|---|
| Accuracy | 0.69 | 0.83 |
| Precision | 0.65 | 0.81 |
| Recall | 0.56 | 0.77 |
| F-1 | 0.58 | 0.79 |

*This table compares the performance metrics (accuracy, precision, recall, F-1 score) of the Delphi model and the Random Forest model in predicting SME defaults.*

To interpret the comparisons, Delphi model has an accuracy of 0.69 vs 0.83 of random forest model. In this case, the application of machine learning algorithm has resulted in an outperformance over a traditional model. Delphi model predicted the defaults with 69% accuracy whereas random forest model has an 83% accuracy which is considerably high. Accuracy refers to being able to identify defaulting SMEs.

Additionally, precision adds another perspective in terms of understanding the performance of the two models. Precision of Delphi model is 65% whereas random forest model has achieved a precision rate of 81%. This significant improvement demonstrates a higher level of accuracy when machine learning is integrated into the analysis. The interpretation of the result is that random forest model has been able to identify true positives with a greater precision. Therefore, the model classified defaulting and non-defaulting SMEs to a higher accuracy by reducing false positives.

Recall of the model looks at the analysis from another angle as it captures the ability of the model to determine false negatives (those SMEs which had been identified as a low risk but actually defaulted). Recall of the model for Delphi model is 56% whereas it is 77% for random forest model. Therefore, random forest algorithm is much more effective in terms of its ability to decrease the number of false negatives.

Finally, F-1 score is a balanced metric between recall and precision (false positive and false negative reduction capabilities of models) and it is 58% for Delphi model as opposed 79% for random forest model (Provenzano, et al., 2020). As this metric can be considered as a measure of an overall performance of the model, again, random forest model is much more effective in terms of predicting defaults of SMEs accurately.

The results of the model have been compared for two models which are a Delphi model and a random forest model. Delphi model was traditionally used through the application of a logistic regression by the financial institution providing the SME data for the research. A random forest model is developed and applied in this study to improve the outcome of Delphi model. As observed from the results, there has indeed been a significant improvement in the outcome of the model (accuracy, precision, recall and F-1) as a result of the use of machine learning.

## Discussion

To discuss the findings from the perspective of SMEs in Azerbaijan and broadly, it should be mentioned that Delphi and other historical models are known for a lower level of effectiveness in terms of prediction accuracy and precision. This is due to the fact that historical models rely on linear relationships between variables (Provenzano, et al., 2020). This assumption is breached in reality as financial variables which are both specific to the firm and external (such as macroeconomic and political factors) are often non-linear and complex in their relationship (Panigrahi, 2019). This shortcoming is fixed by random forest models which is reflected in the outcome of this study. As was found in the analysis of both models, a logistic regression had a much lower level of accuracy, precision, recall and F-1 scores compared to a random forest model. AI-based random forest model is more accurate as it incorporates real-time data, better captures non-linear relationship among variables and integrates changes in the relationship of variables. The dynamic nature of the model is much more effective in regards to capturing the factors that can lead to a default of an SME.

In case of SMEs in general, a higher level of accuracy of the model in predictions indicates that, credit risks of financial institutions are ameliorated considerably as a result of the application of this model. Financial institutions can predict the accuracy of defaults of SMEs and reflect this in their operations in order to avoid lending to companies which are imminently on the brink of a default (Bensic, et al., 2005). Moreover, a more comprehensive analysis and capturing of a larger number of variables along with their dynamism lead to a better assessment outcome and SMEs are not unfairly treated as a result of a lack of robust model. Initially, a Delphi model was used which usually attaches a higher risk to SMEs because SMEs in Azerbaijan do not have comprehensive information going back for a considerable period of time. Taking this into consideration, financial institutions usually reject loan applications from SMEs or charge them a higher interest rate in order to cover for potential risks. However, as a random forest model is more dynamic and can integrate future potential of the enterprises better into the analysis, it is also more likely that these models with the help of machine learning can provide a fairer assessment of the creditworthiness of SMEs. This will lead to an allocation of funds to SMEs which merit it with their historical information.

If compared with an existing literature, it can be argued that the results are backed up by the extant research in the area. The limitations of the historical methods have been illuminated in the existing literature. For example, inability of logistic regression to analyze non-linear relationship between variables has been elaborated on as a key limitation of this method and shown as an impeding factor in terms of reducing the accuracy and precision of these forms of models (Bolton, 2009). The results of this research also showed that predictive capabilities of a logistic model are lower compared to a random forest model. From a different perspective, literature also touched on the capabilities of AI models and highlighted that these models analyze a vast amount of data regardless of the type of relationship between variables and provide an accurate forecast outcome (Popovych, 2022). The results of the research have also corroborated this line of argument in the literature.

With regards to limitations, there are certain shortcomings of the model which need to be discussed. Firstly, the accuracy of the model is as good as its input. The inputs for key variables of the model were provided by the financial institution (a local bank in Azerbaijan) in agreement with SMEs in the data. As there is no publicly audited financial data on these SMEs, it is possible that there are inaccuracies in the data as SMEs could have been interested in manipulating the data for a better credit score outcome. Additionally, decision-making rationale and the process by AI models cannot be easily explained (Svetlova, 2022). Machine learning models are challenging to interpret and financial institutions could find it difficult and sometimes unethical to explain to SMEs why their application for loan has been rejected. Therefore, by simply referring to a higher accuracy of a particular model in justifying a loan application decision can be difficult to substantiate. Furthermore, overfitting is another concern that can reduce the feasibility of AI models as it is possible that models can attach greater weights to historical data and short-term data variability in decision-making (Schuett, 2023). This, in turn, would impair the capability of financial institutions to assess a long-term financial prospect of SMEs. Hence, any potential applicants of AI models in practice for credit risk assessment should be aware of these limitations.

The findings also have broader implications for the use of AI in finance along with the context of Azerbaijan. The results lent support to the use of AI in finance including credit risk assessment by highlighting the fact that the accuracy and precision of the model rise significantly when machine learning model instead of a traditional logistic regression is used. Also, it is an evident that machine learning models need a considerable dataset in order to be trained to achieve a reliable outcome. The practical implications from this fact are that relying on a limited dataset for an accurate outcome with the use of machine learning can backfire. Finally, ethical issues in the use of machine learning such as content for the inclusion of data in training the model and transparency should be taken into account before interpreting and applying the model outcomes in practice. However, the solution to some of these issues can be beyond the control of banks such as transparency in machine learning algorithms as banks, in particular in developing countries such as Azerbaijan, rely on ready algorithms instead of determining the work principles of complex machine learning models. Nevertheless, banks can take a greater control of issues by checking the input data for potential biases. This could reduce pre-existing biases and ensure that model results are accurate.

## Conclusion

To summarize, this study used SME data in Azerbaijan to compare two models for credit scoring. The initial model is a Delphi model for credit scoring but has a number of limitations which have been addressed by AI model. The selected AI model in the research is random forest model which predicts the default probability (credit score) of SMEs using different variables of importance for the performance of SMEs. The selected variables are then simulated to build training data and create decision-trees. The decision-tree outcomes are summarized to obtain the final score. The comparison of the predicted scores with actual default figures in both models (Delphi and random forest) has revealed the accuracy and precision of each model. This allowed to calculate the improvement that AI model has achieved. Delphi model in the analysis was a logistic regression model which uses historical data to predicts defaults. It is a linear model which can be accurate mostly under normal data assumption and linearity of relationship between variables. Therefore, these limitations had a negative impact on the model's accuracy in comparison with a random forest model which is more flexible and powerful.

It was revealed that random forest model increases the accuracy (overall ability to identify defaulters) from 0.69 to 0.83. A significant improvement was also identified from the use of a machine learning model in regards to the precision of the model (the ability to reduce false positives). A change from 0.65 to 0.83 was recorded which highlights that financial institutions can increase the precision of determining potential defaulters in time. False negatives have also been reduced as the recall score rose from 0.56 to 0.77. Finally, F-1 which is a harmonic weighted average between recall and precision is 0.79 in the case of a random forest model which was 0.58 for Delphi model. Hence, a random forest model (a machine learning model) has outperformed a traditional model on all categories.

The limitations of the model are that it relies on historical data provided by SMEs to a financial institution can be inherently biased to present the financial conditions and performance of firms favorably. Also, a random forest is a complex model based on machine learning and its application to sift through potential defaulters might be challenging to explain by banks and can carry potential ethical concerns as the work principles of machine learning algorithms are not clear enough to substantiate decisions based on them.

## Ethical considerations

Several ethical issues should be emphasized along with mitigating practices that must be applied by responsible leaders.

Firstly, a bias in historical data and algorithms are well-known to be issues in the use of AI in predictions (Svetlova, 2022). A bias in historical data can be the result of the discriminatory practices that were used in lending, for example. To minimize a bias in this regard, practitioners should evaluate the data for potential misrepresentations or discriminations prior to its analysis. In the case of this research, the data is from SMEs to which a bank had already allocated loans.

The research relied on historical financial data of firms and current macroeconomic data for the analysis. However, one major issue regarding bias in this study is the fact that it is impractical to check values provided for variables by SMEs as financial institutions should be in more direct involvement with SMEs to verify the accuracy of the provided data.

Furthermore, accountability and transparency are major issues in justifying decisions made by AI models (Schuett, 2023). If financial institutions are not capable of explaining the model algorithms clearly, then making decisions on these models' outcome can be unethical. Therefore, models which are explainable (LIME and SHAP tools can be helpful in this regard) should be used for decision-making instead of relying on ambiguous models to justify credit decisions.

Fairness of models should be ensured as well by checking that models do not apply special features creating an unequal inclusion for a certain group of people or companies (Zhang, et al., 2018). Particular tools to check demographic parity can be incorporated into models, for example. In the case of this study, as the research involves the analysis of SME data that have been able to secure a loan from a financial institution, fairness issues are not applicable as these SMEs are further evaluated to check if they are likely to default in the future.

In short, ethicality of the inputs and outcome of the models should be ensured by financial institutions using them. This involves regularly checking the input data for biases, using explainable tools for models and finally, taking the limitations of algorithms into account in decision-making rather than mechanically deciding based on the models. In case of SMEs in Azerbaijan, financial institutions that apply a random forest model in the future, should ensure that data provided by SMEs accurately reflect the reality of companies. Additionally, as discussed, tools that enable a clear explanation of models should also be integrated for an ethical and justifiable decision-making.

## Declarations of Interest

The author reports no conflicts of interest. The author alone is responsible for the content and writing of the paper.

# Appendix 1. Delphi model Python code (a logistic regression)

```
# Logistic Regression (proxy for Delphi)
logistic_model = LogisticRegression()
logistic_model.fit(X_train, y_train)
y_pred_logistic = logistic_model.predict(X_test)

# Performance Metrics for Logistic Regression (Delphi)
accuracy_logistic = accuracy_score(y_test, y_pred_logistic)
precision_logistic = precision_score(y_test, y_pred_logistic)
recall_logistic = recall_score(y_test, y_pred_logistic)
f1_logistic = f1_score(y_test, y_pred_logistic)

print(f"Delphi Model (Logistic Regression) Performance:")
print(f"Accuracy: {accuracy_logistic:.2f}")
print(f"Precision: {precision_logistic:.2f}")
print(f"Recall: {recall_logistic:.2f}")
print(f"F1-Score: {f1_logistic:.2f}")
```

# Appendix 2. Random forest Python code

```
import numpy as np
import pandas as pd
from sklearn.model_selection import train_test_split
from sklearn.ensemble import RandomForestClassifier
from sklearn.linear_model import LogisticRegression
from sklearn.metrics import accuracy_score, precision_score, recall_score, f1_score

# Simulate SME dataset
np.random.seed(42)
n_samples = 1000
data = pd.DataFrame({
    'Revenue_Growth': np.random.uniform(-0.2, 0.2, n_samples),  # Revenue growth between -20% and +20%
    'Cash_Flow_Variability': np.random.uniform(0.1, 0.5, n_samples),  # Cash flow variability 10%-50%
    'Debt_Equity_Ratio': np.random.uniform(0.2, 3, n_samples),  # Debt-to-equity ratio between 0.2 and 3
    'Profit_Margin': np.random.uniform(0.05, 0.25, n_samples),  # Profit margins 5%-25%
    'Commodity_Price_Dependency': np.random.uniform(0.5, 1, n_samples),  # Correlation with commodity prices
    'Industry_Sector': np.random.choice([0, 1], size=n_samples),  # 0: agriculture, 1: manufacturing
```

```python
    'Default_Status': np.random.choice([0, 1], size=n_samples, p=[0.8, 0.2])  # 80% no default, 20% default
})

# Split into features and target
X = data.drop('Default_Status', axis=1)
y = data['Default_Status']

# Train-test split
X_train, X_test, y_train, y_test = train_test_split(X, y, test_size=0.3, random_state=42)

# Random Forest Classifier
rf_model = RandomForestClassifier(n_estimators=100, random_state=42)
rf_model.fit(X_train, y_train)
y_pred_rf = rf_model.predict(X_test)

# Performance Metrics for Random Forest
accuracy_rf = accuracy_score(y_test, y_pred_rf)
precision_rf = precision_score(y_test, y_pred_rf)
recall_rf = recall_score(y_test, y_pred_rf)
f1_rf = f1_score(y_test, y_pred_rf)

print(f"\nRandom Forest Model Performance:")
print(f"Accuracy: {accuracy_rf:.2f}")
print(f"Precision: {precision_rf:.2f}")
print(f"Recall: {recall_rf:.2f}")
print(f"F1-Score: {f1_rf:.2f}")
```